\documentclass[%
 reprint,
showpacs,
 amsmath,amssymb,
 aps,
 url,
 prd,
 floatfix,longbibliography,
]{revtex4-1}

\usepackage{graphicx}
\usepackage{color}
\usepackage{dcolumn}
\usepackage{bm}
\usepackage{hyperref}

\begin{document}

\title{Observation of Reactor Antineutrinos with a Rapidly-Deployable Surface-Level Detector}

\author{Alireza Haghighat}
 \email{haghigha@vt.edu}
\author{Patrick Huber}
 \email{pahuber@vt.edu}
\author{Shengchao Li}
 \email{scli@vt.edu}
\author{Jonathan M. Link}
 \email{jmlink@vt.edu}
 \author{Camillo Mariani}
 \email{camillo@vt.edu}
\author{Jaewon Park}
 \email{jaewon.park@vt.edu}
\author{Tulasi Subedi}
 \email{tpsubedi@vt.edu}
\affiliation{%
 Center for Neutrino Physics \\
 Department of Physics \\
 Virginia Tech, Blacksburg, VA
}%

\date{\today}

\begin{abstract}
We deployed a small, 80\,kg, antineutrino detector based on solid
plastic scintillator, called MiniCHANDLER for nearly three months at a
distance of 25\,m from a $2.9$\,GW thermal power reactor core at the
North Anna Nuclear Generating Station.  We report the detection of an
antineutrino signal resulting from inverse beta decay at $5.5\,\sigma$
significance with no overburden and minimal shielding. This result
also demonstrates that 3D segmentation can be used to significantly
improve the signal to noise ratio, in this case by a factor of 4.  In
addition, this measurement represents an observation of the positron
spectrum in a small, surface-deployed detector; this
observation of reactor antineutrinos was achieved with a 
mobile neutrino detector mounted in an ordinary, small trailer.
\end{abstract}

\maketitle

\section{\label{Intro}Introduction}

Nuclear reactors have long been known to be a copious source of
electron antineutrinos which are emitted as a
byproduct of nuclear fission.  It is not surprising, therefore, that
neutrinos were proposed as a method to monitor nuclear reactor
operations more than 40~years ago~\cite{Borovoi:1978}.  Neutrino
reactor monitoring is non-intrusive, since it can be performed from
outside the reactor building.  The reactor neutrino signal depends on
both the reactor power and the composition of the reactor core.  In
particular, a core that is rich in plutonium will produce a neutrino
spectrum of lower average energy than a reactor that is rich in
uranium~\cite{Huber:2011wv}. These two signatures can be effectively disentangled by
simultaneously measuring the neutrino rate and energy spectrum.  Case
studies~\cite{Christensen:2013eza, Christensen:2014pva} have revealed
an important advantage of neutrino monitoring compared to the usual
non-proliferation safeguards, which rely on a continuous history of
reactor operations and re-fuelings:  Should this continuity of
knowledge be lost for a reactor, it is extremely difficult to restore.
Neutrino reactor monitoring would not rely on a detailed knowledge of
the reactor's operational history, and thus the continuity of knowledge
issue is avoided altogether.

There are a number of detailed case studies in the literature
highlighting specific potential applications of small above-ground
detectors.  These applications include reactor power
monitoring~\cite{Bernstein:2001cz,Bernstein:2008tj}, monitoring of
spent nuclear fuel~\cite{Brdar:2016swo}, plutonium disposition and
mixed-oxide fuel
usage~\cite{Erickson:2016sdm,Jaffke:2016xdt}. Recently, a detailed
study has been presented ~\cite{Sills649,Carr:2018tak}, how these
capabilities can be applied in a future agreement to denuclearize the
Korean peninsula.

Reines and Cowan used a reactor as the source for their 1956 neutrino
discovery experiment~\cite{Cowan:1992xc}, and since then, many
generations of reactor neutrino experiments have followed, with a
reliance on overburden to shield cosmic rays being an element common
to all.  There have also been a number of successful
safeguards-oriented reactor
experiments~\cite{Korovkin:1988,Klimov:1990,Klimov:1994,Bowden:2006hu,Boireau:2015dda}
starting in the mid 1980s.  Here again they all relied on significant
overburden.  For real-world applications, such as nuclear
non-proliferation safeguards, it is exactly this dependence on
overburden that has prevented the adoption of neutrino technologies.
Practical applications require detectors which can operate with
minimal shielding.  In this paper we describe such a detector
technology and report on the observation of reactor neutrinos in a
small-scale prototype, for other similar measurements
see~\cite{Oguri:2014gta,Carroll:2018kad}. Here we demonstrate a
detector technology which can operate with minimal shielding, has a
small detector volume, and requires no liquid scintillator; in
particular, the neutrino spectrum is measured over a broad range of
energies, including low energies, with high efficiency and good
precision. This combination of features has been previously identified
as crucial towards real-world application of neutrino reactor
monitoring~\cite{iaea}.

In the typical reactor neutrino detector, electron antineutrinos are
observed via the inverse beta decay process (IBD), in which the
neutrino interacts with a hydrogen nucleus in an organic scintillator
producing a positron and a neutron
\begin{equation}
  \bar{\nu}_e + p \to e^+ + n\,.
\end{equation}
The positron deposits its kinetic energy in the scintillator and
annihilates, resulting in a prompt (or primary) flash of light, while
the neutron thermalizes and is captured by a nucleus, producing a
delayed (or secondary) signal.  The signature of the IBD interaction
is the coincidence, in space and time, of positron-like and neutron-like
events.  This compares favorably to the two largest backgrounds which
are 1) fast neutrons from the cosmic ray flux that recoil off of a
proton in the scintillator and capture, and 2) random coincidence
between unrelated positron and neutron-like events.  The random
coincident events have no correlation in space or time, while the fast
neutron events generally share the temporal correlation of the IBD
events but have a larger mean spatial separation due to the greater
initial neutron energy and hence speed.
\begin{figure*}[t]
\includegraphics[width=0.49\textwidth]{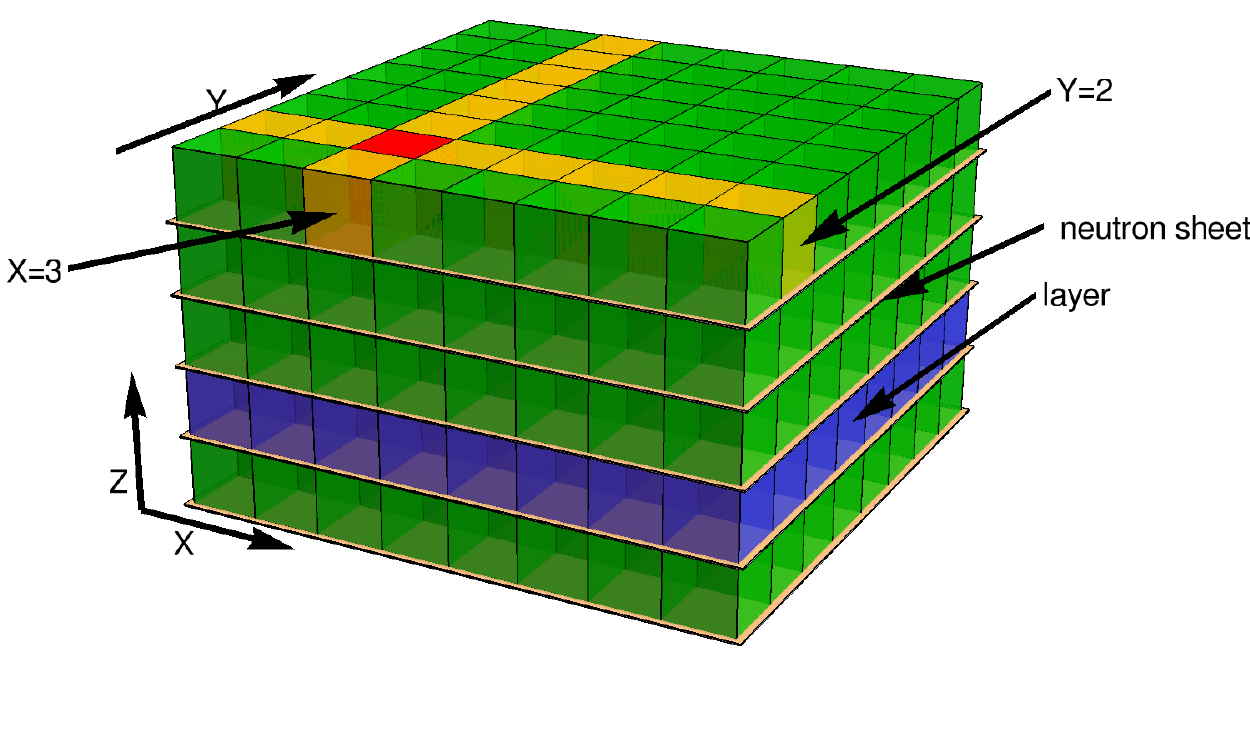}\hspace*{2ex}
\includegraphics[width=0.4\textwidth]{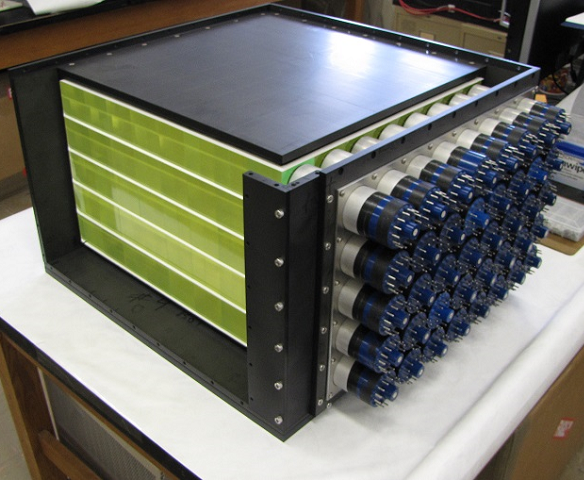}
\caption{\label{Photos} Left: Schematic of the MiniCHANDLER detector
  with the top neutron sheet and PMTs not shown.  Right: the
  MiniCHANDLER detector during assembly, with one side open showing
  the alternating layers of wavelength shifting, plastic scintillator
  cubes and neutron sheets.  }
\end{figure*}

The CHANDLER (Carbon Hydrogen Anti-Neutrino Detector with a Lithium
Enhanced Raghavan optical lattice) detector technology is designed for
the detection and precision spectral measurement of reactor electron
antineutrinos in the high-background surface-level environment.  It
also allows for portable detectors, which are easy to assemble and
easy to maintain, while eliminating the complications and hazards
associated with liquid scintillator.  The CHANDLER design is based on
the optical lattice, which was invented by Raju Raghavan as a part of
the LENS R\&D program~\cite{Grieb:2006mp}.  The Raghavan optical
lattice (ROL) transports light by total internal reflection along rows
and columns of cubes as shown in yellow in the left hand panel of
Fig~\ref{Photos}. The red cube represents the location of the original
energy deposition in this example.  This gives the detector spatial
resolution at the level of a single cube, while at the same time
maximizing the light collection efficiency.  In CHANDLER, the ROL is
formed out of layers of plastic scintillating cubes with a size of
6.2\,cm in a tightly packed rectangular array. These in turn are
stacked in alternating layers with thin neutron detection sheets in
between, as shown in Fig.~\ref{Photos}. In each layer light is
transported along the rows and columns of the cube lattice by total
internal reflection off the inner cube surfaces. This allows the
determination the x-y position of the cube where the original energy
deposition occurred (red cube) and the z-position is obtained as well,
since light is largely confined to the layer in which it was
generated. The neutron detection sheets are semi-translucent, {\it i.e.\ } enough of the
light produced in these sheets will propagate into the adjacent cubes.  
However, the light produced in the cubes has only a small
chance of traversing a sheet, resulting light leakage between the 
layers. The plastics used in the detector naturally maintain a thin 
cube-to-cube or cube-to-sheet air gap, which is required for total internal
reflection.  The plastic scintillator cubes are doped with a wavelength
shifting compound so that the light from the neutron sheets can
be absorbed in the cubes, re-emitted, and then transmitted by total
internal reflection.  The key to this pairing of plastic scintillator
with neutron detection sheets is that the scintillator used in the
neutron sheets releases its light much more slowly than the plastic,
and this results in a clean neutron signature.  Pairing neutron sheets
with scintillator cubes was first implemented by the SoLid
Collaboration~\cite{Abreu:2017bpe}, in which optically isolated cubes
are read out by wavelength shifting fibers running along the edge of
rows and columns of cubes.  Replacing the fiber readout with a ROL
allows to increase the photo-cathode coverage and thus increases light
collection. As a result the energy resolution increases while
maintaining the high spatial resolution and clean neutron tag of the
SoLid design.  When combined, these properties have significant
advantages in the rejection of backgrounds that could otherwise
overwhelm the neutrino signal in a surface-level detector.  For other ongoing simulation studies and R\&D for 2D segmented plastic detectors, see for instance Refs.~\cite{Battaglieri:2010zz,Kashyap:2016bbz,Mulmule:2018efw,Kandemir:2018nsx}.  

The neutron detection sheets and plastic scintillator used in CHANDLER
are sold commercially by Eljen Technology as EJ-426 and EJ-260
respectively.  EJ-426 is composed of micro-particles of lithium-6
fluoride ($^6$LiF) mixed with micro-crystals of silver activated zinc
sulfide (ZnS:Ag) scintillator.  Thermal neutrons are captured by a
$^6$Li nucleus, resulting in a $\alpha$-particle and a triton which, due to
their high specific energy loss, deposit their energy very locally in
the ZnS:Ag scintillator. ZnS:Ag has a scintillation decay constant of
about 200\,ns, which is about 20 times longer than the decay time of
the EJ-260 scintillator used in the cubes.  This large difference in
the scintillation light decay times is used to identify the neutron
captures and separate them from signals originating in the plastic
scintillator.  The cube segmentation in CHANDLER makes it possible to
apply an unbiased prompt/delayed spatial separation cut that is well
matched to the typical positron/neutron separation of an IBD event.
Compared to the standard Daya Bay analysis~\citep{An:2015rpe}, which
uses no spatial separation cut, the coincidence volume in CHANDLER is
reduced by a factor of more than 2000.  In addition, this segmentation
can be used to veto fast neutron events with associated proton recoils
in more than one cube, and to tag the 511\,keV gammas from positron
annihilation in an IBD event.  Together these topological selections
have enabled us to identify the IBD events in a surface-level detector
where correlated background events outnumber the true IBD events by
more than 400 to 1.

The MiniCHANDLER detector is a 80\,kg prototype of the full CHANDLER
detector.  MiniCHANDLER was designed to maximize the detector mass
within our limited project budget with a detector that replicates
light transportation from the middle of the envisioned full-scale
detector.  MiniCHANDLER consists of five layers of an $8\!\times\!8$ cube array
read out by PMTs on only one end of each cube row and column, so that
two faces of the detectors are instrumented with PMTs.  MiniCHANDLER
has six neutron sheet layers: above and below each cube layer. The
sheets are optically connected to the cube layers on both sides, see
Fig.~\ref{Photos}.

The PMTs used in MiniCHANDLER are Amperex XP2202s with a
custom-built, resistive-divider base.  The PMTs are operated at
negative high voltage supplied by a CAEN mainframe with each channel
individually tunable.  The PMT signals were read out by a CAEN V1740
waveform digitizer with 62.5\,MHz sample rate, a 12-bit ADC and 64
channels per card.  To ensure high fidelity with this relatively
sparse sample rate, the PMT signals were first passed through a
pre-amplifier to shape the signal with a 25\,ns time constant.  The
V1470 was internally triggered on every instance of a channel at or
above 14\,ADC counts (ADCC) over baseline.  Each trigger led to a
129-sample read out of all channels in the module starting about 35
samples before the trigger.  Two independently-triggered V1470 modules
were used to read out the full detector.

Data from the waveform digitizers was sent to the DAQ computer over an
optical link, where it was processed through a zero-suppression
algorithm to suppress data from channels in which the waveform only
deviated by 12\,ADC counts or fewer from the baseline.  Only after this zero
suppression was the data written to disk as separate files for each
module.  The two data streams were merged off-line using events from
an external strobe trigger (with a rate of slightly less than 1\,Hz)
to continuously synchronize the merging based on the time-stamps from
the modules' internal clocks.  The strobe trigger merging was used to
estimate the DAQ efficiency, which we found to be greater than 99.5\%.

\section{Reactor and Deployment}

The MiniCHANDLER detector, electronics and DAQ computing were loaded
into a 14\,foot trailer, dubbed the Mobile Neutrino Lab, which was
equipped with a carefully designed quiet power supply, wi-fi
connectivity, and air conditioning, allowing for fully remote
operation. On June 15, 2017, after several weeks of commissioning and
testing at Virginia Tech, the trailer was moved to the North Anna
Nuclear Generating Station in Mineral, Virginia.  The North Anna Plant
consists of two pressurized water reactors, each with a licensed
thermal power of
2940\,MW\@~\footnote{\url{https://www.nrc.gov/info-finder/reactors/na2.html}}.
The Mobile Neutrino Lab was deployed next to Reactor 2, at a distance
of about 25\,m from the center of the core.  At this location it was
approximately 90\,m from the core of Reactor 1, which was therefore
responsible for about 7\% of the neutrino interactions in the
detector~\footnote{During the Reactor 2 shutdown, Reactor 1 was
  operating at full power, so, in the reactor-off subtraction that
  will be performed, its events are counted as a part of the
  background.  Therefore, we calculate the expected rate relative to
  Reactor 2 alone.}.  The detector and DAQ were up and running in less
than one day, which marked the start of the site specific
commissioning.  To combat the increased thermal neutron rate from the
reactor, the detector was surrounded by a layer of 1-inch thick
boron-loaded polyethylene with holes for the PMTs.  The natural gamma
rate at the reactor site is higher than the one experienced at
Virginia Tech.  To combat this we added an inch of lead shielding
below the detector, and on the two sides closest to the containment
building.

With commissioning complete, the data run began on August 9, 2017 and
lasted through November 2, 2017.  During this time we took 1133.6 hours
of usable reactor-on data and 675.4 hours of reactor-off data. The data
are divided into eight periods, where the transition between periods
corresponds to changes in the operational state of either the detector
system or the reactor.  Table~\ref{tab:periods} describes the
different periods, and lists the reasons for the start of each new
operational period.  Of particular note is the transition from period
2 to period 3, which corresponds to a shift in the trigger threshold
from 10 ADCC to 14 ADCC\@.  This became necessary when the rate of
low-energy gamma rays increased due to the arrival of shipping 
containers of mildly activated equipment that were parked next to 
the Mobile Neutrino Lab in preparation for the refueling of Reactor 2.

\begin{table}[tb]
    \centering
    \begin{tabular}{c|c|l} 
         
         \textbf{Period} & 
         \textbf{Runs} & \textbf{Reason for New Period} \\ \hline
         1 & 258 & 10 ADC threshold \\ 
         2  & 164 & Streamlined disk I/O \\
         3  & 255 & Change to 14 ADC threshold \\
         4  & 5   & Reactor ramp-down \\
         5  & 569 & Reactor off \\
         6  & 118 & High voltage re-tune \\
         7 & 49  & Reactor ramp-up \\
         8 & 476 & Reactor at full power \\ 
    \end{tabular}
    \caption{Description of the operational run periods. Each run corresponds to 1 hour.}
    \label{tab:periods}
\end{table}

\section{Calibration}
For the study described here, a highly-accurate energy model and
reconstruction was neither a requirement nor an objective.
Nevertheless, matching the known energy dependence of reactor
neutrinos in an observed reactor-on excess was an essential
confirmation of neutrino detection.  In addition, we were motivated to
test a novel energy calibration source made possible by the
high-segmentation of the ROL\@.  Specifically, in polyvinyl toluene, a
minimum ionizing particle has a $dE/dx$ of about
2\,MeV/cm~\footnote{\url{http://pdg.lbl.gov/2018/AtomicNuclearProperties/HTML/polyvinyltoluene.html}},
which means that a muon, passing vertically through a 6\,cm cube
deposits an average energy of around 12\,MeV\@.  In the following
section we describe how we used vertical muons to measure the light
pattern from every cube location in the detector, and how this allowed
us to fix the energy scale at around 12\,MeV\@.  Here, we assume a
proportional energy response for energies below 12\,MeV\@.

The PMT high voltage was initially tuned to align the muon peaks
across all channels to 1500\,ADC counts.  To account for gain fluctuations,
the muon peak was measured in each channel for each run and the
measured ADC values were scaled to realign the muon peaks.  In this
context the muon peaks are not limited to vertical muons, which have
limited statistics in a single run, but include all triggers across
all cube positions.

\section{Event Reconstruction}

Neutron identification in MiniCHANDLER is based on pulse shape
discrimination, using the factor of 20 difference in the scintillation
light decay times between the neutron sheets and the scintillator
cubes.  A naive particle identification (PID) variable can be formed
from the ratio of the area under the waveform to its peak value.
Large values of this variable correspond to neutron-like events, while
small values correspond to signals generated in the plastic
scintillator.  Large signals, with peak values greater than 1000\,ADCC
were eliminated from considerations.  If a signal satisfies our neutron
PID criterion in at least one PMT channel, the whole event becomes a neutron
candidate.

Instrumental effects in MiniCHANDLER, such as PMT flashers and analog
overshoot from an earlier large pulse, can generate signals that
satisfy this naive neutron PID selection, fortunately these effects
almost never replicate the decaying light pattern of an energy
deposition in the neutron sheets.  We used a template-based
$\chi^2$-criterion to reject these instrumental backgrounds from the
neutron candidate list.  To obtain the neutron-template we started with
a sample of 100 hand-selected neutron capture waveforms.  Each
waveform was divided into eight regions.  In each region, the ADC
counts over baseline were summed, and these sums were divided by the
total over all regions to form normalized amplitudes.  Then these
normalized amplitudes were averaged over the 100 hand-selected
waveforms to form the neutron-template.  Since, events in the plastic
scintillator have short pulses which are contained entirely in the
first region, the gamma-template is trivial.  With these templates the
neutron selection proceeds as follows.

Within each view of each layer, we select the channel with the
highest amplitude signal, compute its normalized amplitudes and
uncertainties in the eight regions, and compute the $\chi^2$s relative
to both the neutron- ($\chi^2_n$) and gamma- ($\chi^2_{\gamma}$)
templates.  The reduced $\chi^2$s from both the $x$- and $y$-views are
summed and we select good neutrons satisfying the criteria
$\sum_{x,y}\chi^2_{n_i}/\nu_i<8$ and
$\sum_{x,y}\chi^2_{\gamma_i}/\nu_i>150$, where $\nu_i$ is the number
of time bins in the template, effectively the number of degrees of
freedom. This $xy$-matching fixes the position of the neutron
candidates.

Once neutron identification in a layer is done, we check the
consistency of $xy$-matched neutrons from different layers.  Neutron
candidate events generally exhibit low occupancy in the detector.
Therefore, the $xy$-position of an event is simply given by the
location of those PMTs which see the most light.  For about half of
all neutron capture events we see light on only one side of the
neutron sheet.  We call these events ``cube" neutrons since we can not
distinguish whether the capture happened in the sheet above or below
the cube.  In these events the neutron $z$-position is assigned to the
middle of the cube.  For the remainder of events the neutron capture
is seen on both sides, and the neutron capture position is known at
the sheet level; we call these events ``sheet" neutrons.  Any event
with more than one neutron candidate among the 5 layers is rejected.
Tests with Li-free neutron sheets in our MicroCHANDLER prototype have
shown that in the absence of $^6$Li there are practically no
neutron-like signals in the detector.  Therefore, for the purpose of
this analysis, we can treat all neutron-like events as neutrons without
introducing any bias.

Event reconstruction for prompt events is somewhat more complicated
than for neutrons because the number of active cubes in the detector
is often greater than one.  This is due to the Compton scatter of positron
annihilation gammas in IBD events, and to the possibility of multiple
proton recoils in fast neutron backgrounds.  In order to use this
topological information, we need a reconstruction that is capable of
evaluating energy depositions in multiple active cubes. Here a
challenge arises when there is more than one active cube in a single
detector layer.  This is a non-trivial problem, because in each
detector layer we have $2\times8$ observed PMT signals, but there are
$8\times 8$ unique cube locations.  If we knew the true energy 
deposition in each cube in a layer, expressed as a 64-component vector, 
$\mathbf{e}$, then we could write an expression for expected PMT 
responses as the 16-component vector, $\mathbf{p}$.  This forward 
problem is represented by
\begin{equation}
\label{eq:transfer}
    \mathbf{p}=\mathbf{M}\cdot\mathbf{e}\,,
\end{equation}
where $\mathbf{M}$ is the $16\times64$ transfer matrix.  Each element
of the transfer matrix, $M_{ij}$, describes the size of the signal in
PMT $j$ arising from a 1\,MeV energy deposition in cube $i$.  This
transfer matrix includes all effects arising from light propagation,
including attenuation and scattering, and the electronics cross talk.
Although, about 80\% of the light detected is observed in the PMTs at
the ends of the row and column centered on the emitting cube, the
remaining 20\% of light is spread out across the other PMTs in the
plane.  This unchanneled light is due to tiny imperfections in the
ROL, and to scattering in the bulk of the plastic cubes. In addition 
to the unchanneled light, there is a bi-polar, inductive-pickup 
cross talk which is observed in channels neighboring one with a 
large amplitude pulse.

Our objective is to invert this matrix equation to solve for $\mathbf{e}$, the 
vector of cube energies, but first, we had to determine the transfer matrix, 
$\mathbf{M}$.  Even then, there is no exact solution to 
Eq.~\ref{eq:transfer}, since $\mathbf{M}$ has no inverse.

A data driven approach was used to determine the elements of the
transfer matrix.  This is the best way ensure that all effects are
properly accounted for.  We used vertical muons, which are easily
identified in our detector by requiring that the observed light be
consistent with coming from the same single cube position in each
plane.  By definition, a vertical muon produces light in only one cube
per plane, and that cube's position is well-known from the $xy$-coordinates of
the vertical muon.  Unchanneled light and electronic cross talk spread
this signal over all channels in the plane.  By collecting a large
sample of vertical muons, which occur at a rate of 0.7\,Hz across the
detector, we measured the response of each PMT in each plane to the
energy depositions from every cube position in that plane.  According
to our simulation, the most probable energy deposition for a muon that
satisfies the vertical selection is 11.42\,MeV/cube.  The transfer
matrix elements were scaled to an equivalent energy of 1\,MeV\@.  In
constructing the final transfer matrix, which is applied to all
layers, we average the elements from the matrices measured in just the
middle three detector layers, and we did this because we can only be
certain that a ``vertical" muon's path through a layer was fully
contained in a single cube when there are confirming hits above and
below that layer.  In the case of the top and bottom layers one of
these confirming hits is missing.

\begin{figure}[tb]
\centerline{\includegraphics[width=0.48\textwidth]{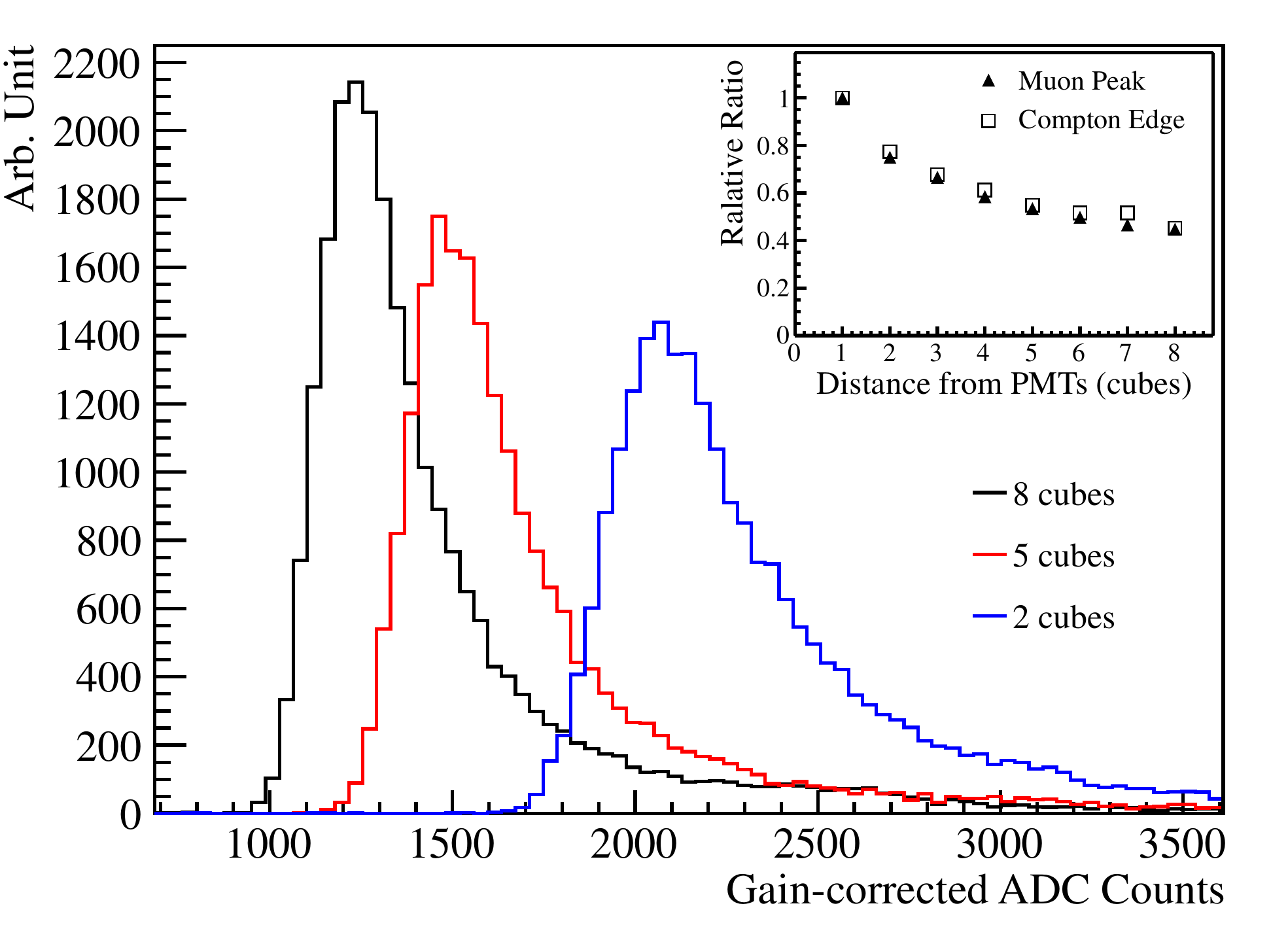}}
\caption{\label{vertical_muons}Shown are vertical muon energy spectra at three
  distances from the PMTs.  The inset shows the effective
  light attenuation in the ROL as determined from vertical muons, at
  11.42\,MeV, compared to the attenuation measured with the
  Compton edge of the 1.275\,MeV gamma from $^{22}$Na at around 1\,MeV\@.  }
\end{figure}

A sample of the vertical muon spectra from cubes at three different
distances from the PMTs is shown in Fig.~\ref{vertical_muons}.  The
width of these energy distributions comes from of the natural
Landau distribution in $dE/dx$, the geometrical acceptance for muons
which are not perfectly vertical, and the intrinsic resolution of the
detector.  The shift in the peak position, as a function of distance
from the PMT, shows the spatial dependence of the detector 
response function.
This effect is explicitly accounted for in the transfer matrix.  
Throughout the region of interest, this spatial dependence 
is independent of the deposited energy.  This is illustrated in the inset of 
Fig.~\ref{vertical_muons}, which compares the effective attenuation curve 
for vertical muons to the one derived from the $\sim\!1$\,MeV Compton edge of 
1273~keV gamma from a $^{22}$Na source.

Both the unchanneled light and the electronics cross talk scale with
the amount of light detected in the primary channel, but the variances
of the unchanneled light and electronics cross talk components do not.
For unchanneled light the variance scales with the Poisson statistics
of the photons at the PMT cathode, while for cross talk the variance
scales with the electrons at the PMT anode.  The future, full detector
will use electronics without cross talk.

GEANT4~\cite{Dellacqua:1994iv} was used to compute the true
cube-by-cube energy depositions for a set of simulated gamma and IBD
events.  Using the transfer matrix, $\mathbf{M}$, and random
fluctuations drawn from a Poisson-distribution with the appropriately
scaled variances, this truth information was propagated to create a
Monte Carlo realization of the measured PMT signals.  This sample was
used to test and tune the event reconstruction.  As previously stated,
Eq.~\ref{eq:transfer} has no exact solution, any approximate solution
must compensate for the lack of observables by some regularization
scheme.  This essentially amounts to using a Bayesian prior to select
among the possible solutions.  For our analysis the goal is to
correctly reconstruct the number of cubes with a non-zero energy
deposition, with a preference for suitable solutions with the fewest
active cubes.  This matches our expectation for IBD events, which the
Monte Carlo has shown will almost never have more than five cubes with
true energy depositions above the detection threshold in the
MiniCHANDLER detector.

Using the variance found from data we constructed a suitable likelihood
function, $L$, to measure how well a given set of reconstructed cube
energy depositions, $\mathbf{e}_r$, corresponds to the measured PMT
signals, $\mathbf{p}_s$.  To minimize $\hat{L}\equiv-\log L$ we used the
following algorithm:  First set all $e_r(x,y)=0$ ({\it i.e.} all positions 
have an initial energy deposition of zero),such that the set of cubes with 
non-zero energy deposition, $\lambda$, is initially empty.
\begin{enumerate}
    \item Find the additional cube $(x,y)$ that yields the smallest $\hat{L}$, 
      when $\hat{L}$ is minimized by varying $e_r$ for the cubes in set
      $\lambda$ plus the new cube $(x,y)$. 
    \item If $\min(\hat{L}(\lambda))-\min(\hat{L}(\lambda+(x,y))) < L_c$, 
      go to step 5.
    \item Add cube $(x,y)$ to the $\lambda$ set.
    \item While $\lambda$ has less than five cubes, return to step 1.
    \item If $\hat{L}<L_g$, accept event as reconstructed, otherwise
      declare the reconstruction failed. 
\end{enumerate}
This algorithm allows the reconstruction to assign energy depositions
to additional cubes as long as the improvement in $\hat L$ is
sufficiently large ($>L_c$).  This cutoff prevents over-fitting, since
adding a cube always will decrease $\hat{L}$.  The reconstruction is
limited to no more than five cube in a layer, which is a conservative
upper limit relative to the observation in our Monte Carlo that IBD
events have no more than five active cubes in the whole detector.
Step 5 ensures that the fit is a good match to the data by requiring
the final $\hat{L}$ to satisfy a quality criterion ($<L_g$).  This is
rarely violated by IBD events in the Monte Carlo, but in data,
where the true composition of the event types is unknown, we find about
7\% of events fail this criterion in at least one layer.  We thus
quote a reconstruction efficiency of 93\%, but, since the IBD events
generally do not share the typical characteristics of the events that
fail the reconstruction, we presume it to be much higher for true IBD
events.  The fit cutoff, $L_c$, was tuned on Monte Carlo IBD events
and the quality parameter, $L_g$, was tuned on background data
samples. This maximizes the reconstruction fidelity to the true cube
positions and energy depositions, and minimizes reconstruction
failures.

As a test, the reconstruction was applied to a sample of vertical muons
from across the whole detector.  The resulting energy spectrum was
fitted to a convolution of Landau and Gaussian distributions.  The
fitted peak value was in good agreement with the most probable energy
deposition from the simulation.  We interpret the fitted Gaussian
$\sigma$ to be the average energy resolution at 12\,MeV, which was
found to be 2.6\%.  If the resolution scaling is purely stochastic
this would correspond to an average resolution of approximately
10\%$/\sqrt{E\rm(MeV)}$.

\section{IBD Analysis}

To compute the expected IBD spectrum and number of events, we use the
Huber-Mueller reactor flux model~\cite{Huber:2011wv,Mueller:2011nm},
and the IBD cross section from Ref.~\cite{Vogel:1999zy} with a neutron
lifetime of 878.5\,s.  The thermal reactor power is taken to be
2.94\,GW and the core-detector distance is 25\,m.  The detector mass
is 80\,kg, comprising $4\times10^{27}$ target protons.  From
simulation we compute that 46\% of all IBD neutrons in the detector
capture on $^6$Li.  Of these 34\% are lost when we discard the first
40\,$\mu$s in $\Delta t$.  The entire reactor-on data set is comprised
of 1133.6 hours of good data.  Under these assumptions we expect about
3500 Li-tagged, IBD events.  Given the uncertainties in the Monte
Carlo, the reactor distance, and spill-in/spill-out effects, it is
difficult to assign a firm error, but 10-20\% appears reasonable.  IBD neutrons created outside the detector and IBD neutrons
  reflected back into the detector are difficult to simulate precisely
  and Monte Carlo tests indicate that this effect is below 10\%. The
  reactor detector distance is known to within 1\,m, translating into
  about 8\% uncertainty. The reactor power is know within 1\% and fuel
  burn-up has not been corrected for, but this effect does not exceed
  5\%.

  In quadrature these errors would add up to 12-14\%, but for instance
  the distance uncertainty is non-Gaussian, {\it i.e.} the distance
  could be between 24 and 26\,m but it certainly is not 26.5\,m; the
  same holds for the neutron capture efficiency or the burn-up
  effect. So these errors can \emph{not} be added up in
  quadrature. The exact value of the systematic uncertainty on the
  number of expected IBD events has no impact on the statistical
  significance of the IBD signal, since the IDB signal is derived
  entirely from a comparison of reactor-on and reactor-off data
  without recourse to the expected number of IBD events.  The reason
  to calculate the expected number of IBD events is to check
  whether the number of expected events is consistent with the 
  observed events, which it was found to be.  

GEANT4 was used to simulate the cube-level energy depositions from IBD
events, but we did not use it to propagate photons through the ROL.
Instead we generated the PMT signals in ADCC using the forward
transfer matrix derived from vertical muons, followed by a Poisson
smearing based on the observed and scaled variances.  The simulated
PMTs signals were run through the reconstruction and event selection
just like the data.  Therefore, any non-linearity in the reconstructed
energy spectrum should be common to both data and Monte Carlo, at
least to within the precision of this analysis.

\begin{figure}[t]
\includegraphics[width=\columnwidth]{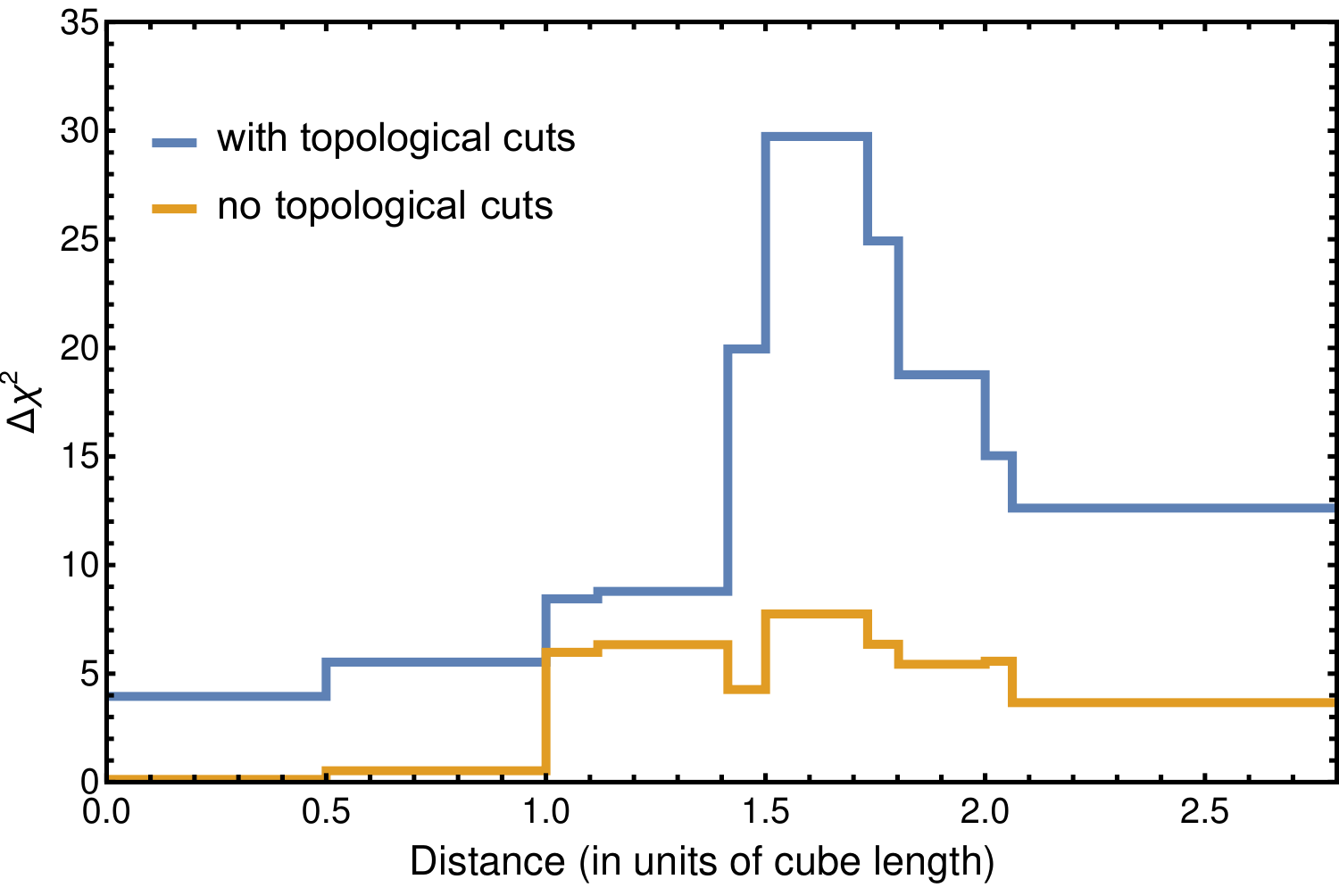}
\caption{\label{distance_cut} The significance of the IBD signal, in
  $\Delta \chi^2$ relative to the null hypothesis, plotted as a
  function of the maximum allowed distance between the cube which has
  the prompt signal and the one which has the delayed  signal,
  in units of cube lengths.  The significance is determined by varying
  the cuts in the data and running through the full analysis.}
\end{figure}

To form IBD event candidates, we begin by matching each neutron
capture candidate with all non-neutron events with a successful
reconstruction from the preceding 1000\,$\mu$s. Next we apply a
prompt/delayed spatial separation cut.  The prompt event position is
assigned to the center of the most energetic cube of the primary
event.  To assign the position of the delayed event we distinguish
sheet and cube neutrons, as explained previously.  As we expand the
allowed separation, more correlated events are included in the sample.
At short distances we find the largest enrichment of true IBD events,
but as the separation grows fast neutron events start to dominate.  To
select the optimal separation cut, we studied IBD signal significance
as a function of the separation cut.  Figure~\ref{distance_cut} shows
the $\Delta \chi^2$ relative to the null hypothesis, plotted as a
function of the maximum allowed prompt/delayed separation.  The
stepped nature of this plot is due to the quantization of separation
distances inherent in our assignment of event positions in the cube 
structure.  The significance peaks at a separation of 1.5 cube 
lengths, or 9.3\,cm.  At this distance the cut includes the 19 nearest 
cube positions, and 20 nearest sheet positions.  From our
IBD Monte Carlo, we estimate that 67.3\% of true IBD neutrons are
captured within this region.  As the fast neutron rejection 
improves in future incarnations of the detector, this cut can be
opened up to improve the IBD efficiency while maintaining maximal
significance.

The 3D segmentation of MiniCHANDLER allows us to further select events
based on the topology of the event.  Under perfect conditions, one
would design cuts to specifically tag the two 511\,keV positron
annihilation gammas.  In the current MiniCHANDLER detector this is not
practical for two reasons:  First, the detector is too small to
efficiently contain the first Compton scatter from both annihilation
gammas.  Second, with the current light collection scheme the
detector's energy threshold is about 50\,keV, and at that level, many
of the annihilation gamma Compton scatters are unseen in the detector.
Therefore, we have implemented a set of cuts to retain events with any
hint of the positron annihilation gammas, while rejecting events
that are clearly inconsistent with their presence.  Specifically, we
required there to be least 1 cube, beyond the primary (or highest
energy) cube, with energy deposition in the range of
$50\,\mathrm{keV}\leq e_r\leq 511$\,keV\@.  Further, we require that
the sum of energies in all cubes, excluding the primary cube and its
most energetic immediate neighbor, be no more than 1022\,keV, and that
outside of the those two cubes there is no single cube energy
above 511\,keV.  These last cuts are designed to remove fast neutrons
with multiple proton recoils.  As can be seen by comparing the blue
and orange lines in Fig.~\ref{distance_cut}, these topological cuts
improve the signal significance from $\Delta \chi^2=7.7$ to $\Delta
\chi^2 = 29.7$, or equivalently the signal-to-noise is improved by a
factor of about 4.  This demonstrates that the fine-grained 3D
segmentation at the core of the CHANDLER technology adds considerable
value relative to the coarser-grained 2D segmentation used in other
contemporary detectors~\cite{Ashenfelter:2015tpm, Allemandou:2018vwb}.
With anticipated improvements to the light collection, and a larger
detector to better contain the annihilation gammas, the efficacy of
these topological cuts should be significantly enhanced.

\begin{figure}[t]
\centerline{\includegraphics[width=0.48\textwidth]{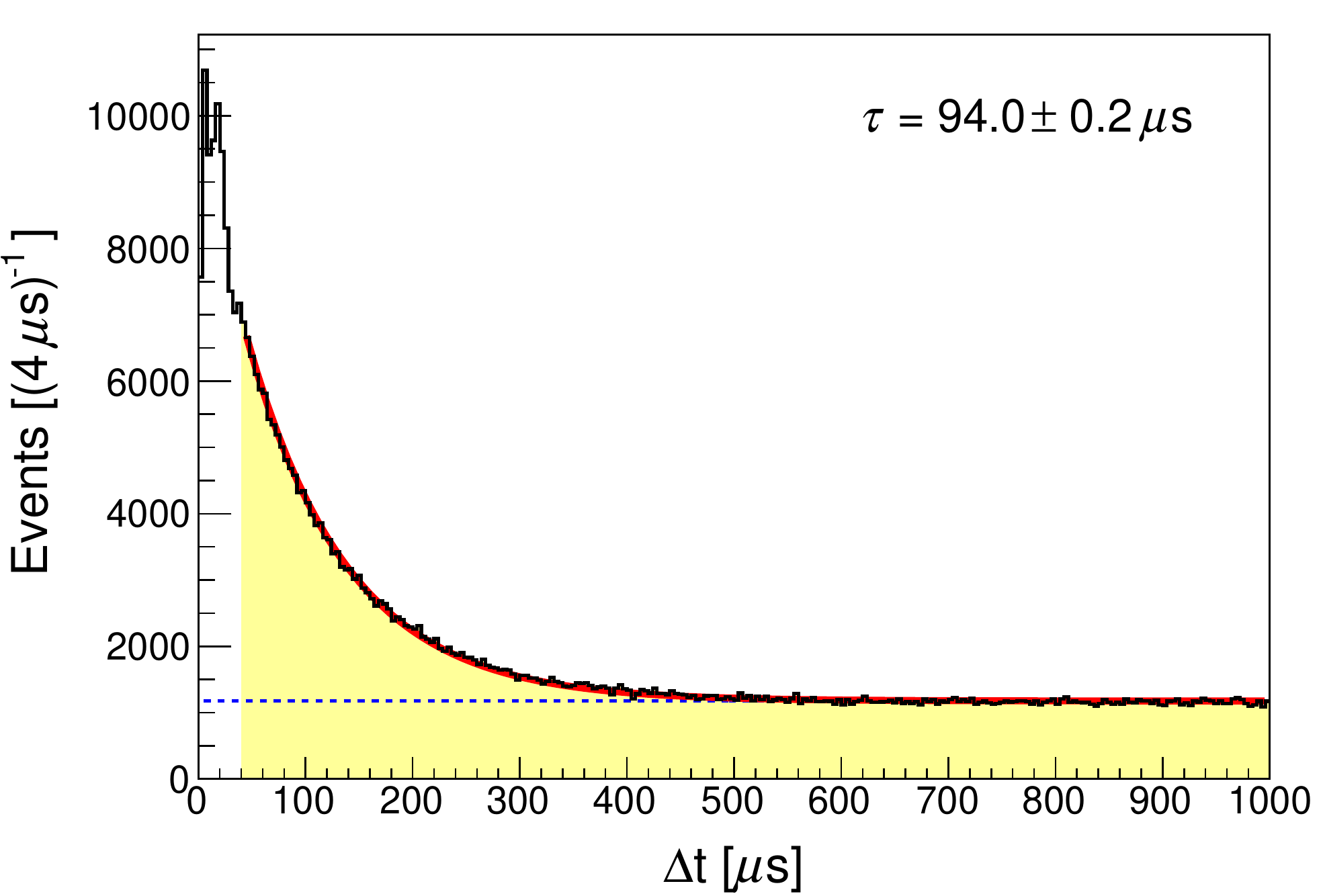}}
\caption{\label{Delta_t} Shown is a histogram of $\Delta t$ for all
  events in the energy range of 0.5--20\,MeV.  This distribution is
  fitted with an exponential plus constant to extract the true
  correlated events.  The fit begins above 40\,$\mu$s (in yellow) to
  bypass the low $\Delta t$-region, where instrumental effects distort
  the distribution.}
\end{figure}

The surviving events are split into reactor-on and reactor-off samples
(see Tab.~\ref{tab:periods}).  In each sample, they are sorted by their
reconstructed prompt energy into 20 bins from $0.5-20$\,MeV, with the
lowest energy bin being 0.5\,MeV wide and all other bins being 1\,MeV
wide.  In each energy bin, the prompt/delayed $\Delta t$-distribution
is fitted with an exponential plus flat function.  The exponential
time constant, $\tau$, is fixed to 94\,$\mu$s, as was determined from
a single $\Delta t$-fit to the data from all energy bins and reactor
periods (see Fig.~\ref{Delta_t}).  These $\Delta t$-fits are used to
statistically separate the time-correlated events (the exponential
component) from the random coincident events (the flat component).
Using all positron candidate events in the 1000\,$\mu$s proceeding a
neutron --- as opposed to just using the first event, or vetoing all
events when two or more positron candidates are observed --- ensures
that the $\Delta t$-distribution from the random coincident
contribution is flat over all times.  Then, by fitting this distribution 
out to more than 10 neutron capture lifetimes, we get a high-fidelity,
high-statistics measure of the random component, which we then subtract
to get the correlated rates.  A sample $\Delta t$-distribution, with 
fit, is shown in Fig.~\ref{Delta_t}.  Due to
effects related to the analog side of our signal processing chain, we
exclude the first 40\,$\mu$s from the fit.  In the subsequent analysis,
this results in a loss of 34\% of all true IBD events.

\begin{figure*}[tb]
\includegraphics[width=\textwidth]{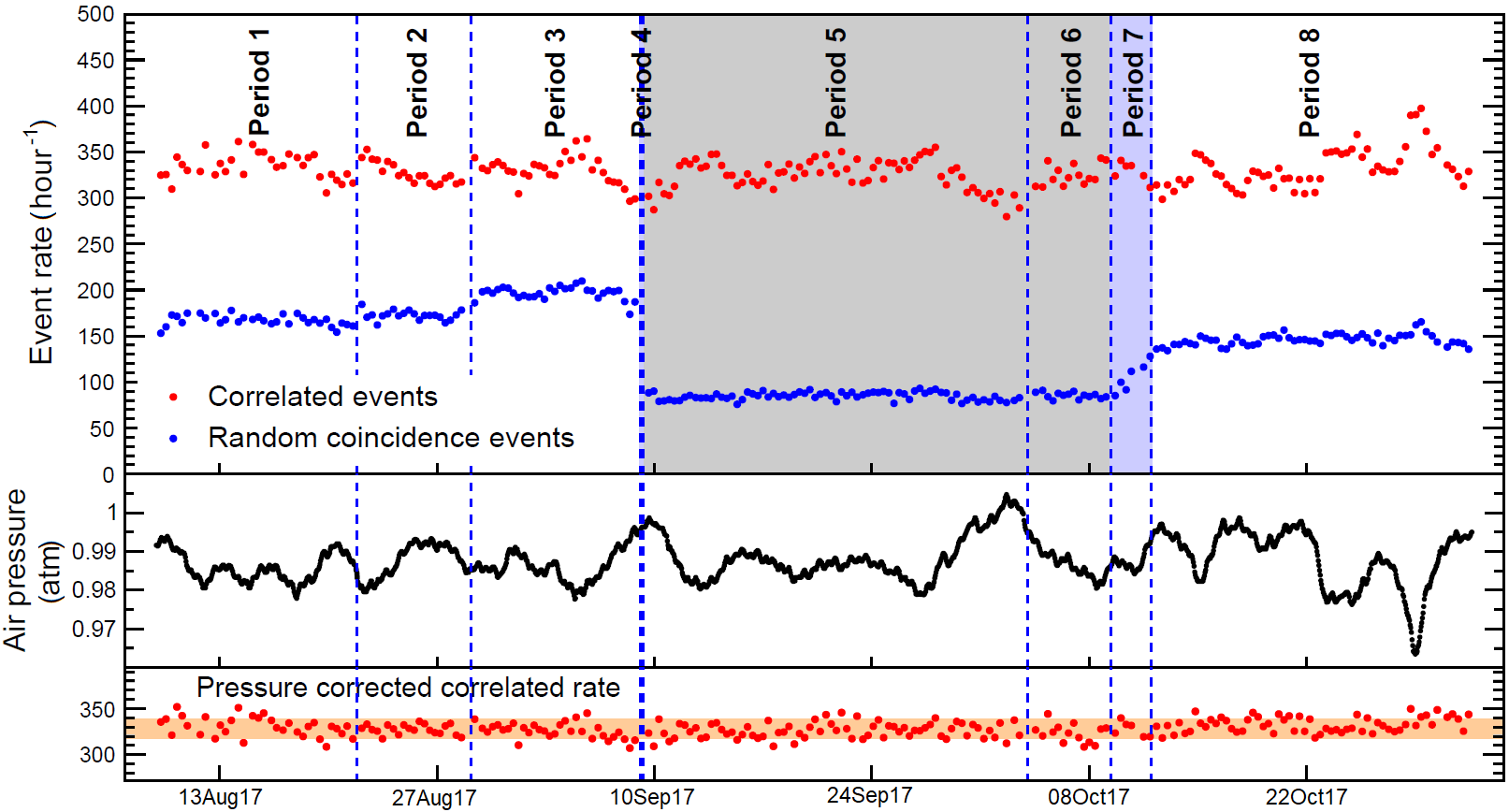}
\caption{\label{Stability} The top panel shows the rates, as a
  function of time, for correlated (red) and random coincident (blue)
  events as extracted from a fit to the $\Delta t$-distribution in
  each 8 hour period.  These events were selected 
  without topological cuts applied in order to 
  enhance the fast neutrons events relative to IBD.  The middle panel
  shows the time dependence of the atmospheric pressure, which is
  anti-correlated with the variations of  the correlated event
  rate.  The bottom panel shows the correlated event rate, corrected
  for atmospheric pressure as described in the text.  The orange band 
  is the average 1\,$\sigma$ uncertainty on the correlated event rate.
  The run periods are described in Table~\ref{tab:periods}.  The gray 
  shaded periods (5 and 6) correspond to reactor-off. The blue shaded 
  periods (4 and 7) correspond to reactor power ramping, and are not 
  used in the IBD analysis.  }
\end{figure*}

In the final step of the analysis, we perform a background subtraction
by taking the difference of correlated events in the reactor-on
periods to those in the reactor-off periods.  In this step there is a
danger of introducing structure into the energy spectrum if the
detector operation was not stable over time.  Fig.~\ref{Stability}
shows the correlated (red) and random coincident (blue) event rates,
as a function of time, as extracted from the $\Delta t$-distribution
fits, but without topological cuts applied, to enhance the fast
neutron events relative to IBD\@.  

The random coincident rate shows
large variations between periods, which are linked to specific
operational events at the plant.  For example, during the shutdown,
when the thermal neutron rate from the reactor was essentially zero,
the random coincident rate was cut in half.  Similarly, at the start
of period 3 we see a slightly higher random coincident rate, which 
corresponds to the arrival of several shipping containers as discussed 
earlier.  The increased the trigger threshold from 10 to 14\,ADCC that 
followed this event was applied after the fact in software to the data 
from periods 1 and 2 to ensure uniformity across the periods.

On the other hand, the period-to-period jumps observed in the random coincident rate are
not seen in the correlated event rate.  Instead, we see smaller 
undulations which are anti-correlated with the atmospheric pressure.
This is exactly what one would expect if the correlated rate was
dominated by fast neutrons in the cosmic ray flux, as should be the
case here.  It is well known that the cosmic neutron rate is related 
to the atmospheric pressure, which is a measure of the mass of the 
atmosphere above.  The air pressure shown in the middle panel of
Fig.~\ref{Stability} was measured at the Louisa County Airport,
located 16.7\,km from the North Anna Nuclear Generating Station, and
was obtained from the NOAA
website~\footnote{\url{https://www.ncdc.noaa.gov/cdo-web/datatools/lcd},
  Station ID: WBAN:03715}.  Using this data, we compute a correction
factor for the measured pressure, $P$, relative to the average
pressure, $P_0$, which is equal to $e^{-\alpha(P-P_0)}$ with
$\alpha=7.3\,\rm{atm}^{-1}$~\cite{doi:10.3402/tellusa.v14i1.9527}.
In the bottom panel of Fig.~\ref{Stability}, this correction factor 
is applied to the measured correlated event rates, which, once corrected, 
are stable across all data taking periods.  The orange band represent the 
average statistical error of the correlated event rate as measured in 8 
hour blocks.

While the air pressure's impact on the fast neutron rate is a
well-understood phenomenon that can be compensated for in the overall
rate, it was not immediately clear whether differences in the average air
pressure between the reactor-on and reactor-off periods could
introduce an energy dependence in the correlated rate that could mimic
an IBD signal.  To test this hypothesis, the reactor-on data were
split evenly into high-pressure and low-pressure sets and the analysis was
run on both halves.  The IBD excess measured in the two sub-samples agreed 
to within $1\,\sigma$.

\begin{figure}[t]
\includegraphics[width=\columnwidth]{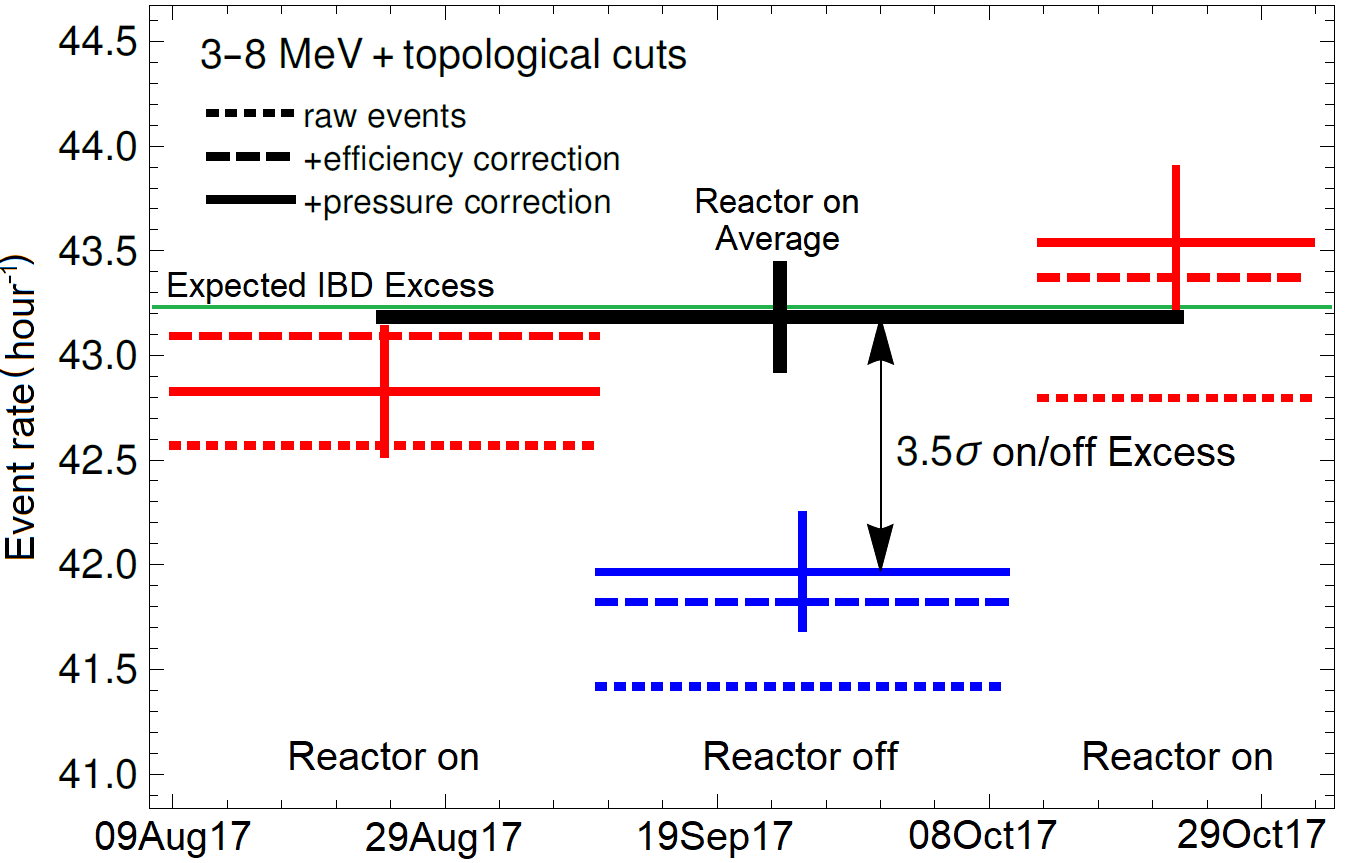}
\caption{\label{fig:total} The results of the total correlated rate 
analysis are consistent with the conclusions of the spectral analysis.}
\end{figure}

An analysis based purely on the total correlated event rates 
was conducted as a cross check of the spectral analysis.  In this case, 
the air pressure correction was applied run-by-run.  This incorrectly 
rescales the IBD signal events, however, since the correction is at most 
5\% in any given run, but this is of little consequence for the current purpose. 
Also, the DAQ live-time efficiency has a slight systematic difference 
between reactor-on and reactor-off runs due to the extra thermal neutron 
triggers when the reactor is on.  We measured this efficiency in each run 
by comparing  the number of recorded strobe triggers to the number that 
were sent.  The statistical significance of this counting analysis is a 
strong function of the signal-to-noise ratio, so  we applied a 3--8\,MeV energy 
cut, which should retain 58\% of IBD events while reducing the 
background by a factor of 3.6.  To further enhance the signal-to-noise ratio 
the topological cuts are applied.  The data is in three time periods: before, 
during and after the reactor shutdown.  Respectively these correspond to the 
data taking periods of Tab.~\ref{tab:periods} as periods 1--3, periods 5--6, and 
period 8\@.  The results are shown in Fig.~\ref{fig:total}.   

This analysis finds an on/off-excess of $1.22\pm0.35$\,events/hour,
which corresponds to a 3.5$\,\sigma$ significance.  This compares well 
to the expected IBD rate in the 3--8\,MeV range of 1.27\,events/hour, 
which demonstrates that the topological cuts are highly efficient for true 
IBD events.  Multiplying the observed excess by the total reactor-on  
time and correcting for the 58\% efficiency of the 3--8\,MeV cut this
corresponds to $2418\pm700$ events, which is entirely consistent with the
result of the spectral analysis.  The lower significance
of the rate only analysis, is partly due to the inefficiency of the 
energy cut and partly to the absence of information from the IBD
spectrum.  If we use the full 0.5--20\,MeV range, we find an on/off
excess of $2.48\pm0.94$ events/hour compared to an expectation of 2.17 IBD
events/hour.  This corresponds to only 2.6$\,\sigma$ significance, 
demonstrating a well-known feature of rate-only analyses: tight cuts
must be applied to obtain a suitable signal-to-noise ratio, resulting in 
a lower overall IBD efficiency.

It is worth noting that the pressure and DAQ efficiency corrections 
are comparable in size to the expected IBD excess in this rate
analysis. Therefore, when the signal-to-background ratio is low,
this type of analysis may be less reliable than desired. Also,
random coincident rates can be strongly correlation with the reactor
power, see Fig.~\ref{Stability}, and therefore must be separated in a
robust way, such as the $\Delta t$ fit method illustrated in
Fig.~\ref{Delta_t}.

\begin{figure*}[t]
\includegraphics[width=0.7\textwidth]{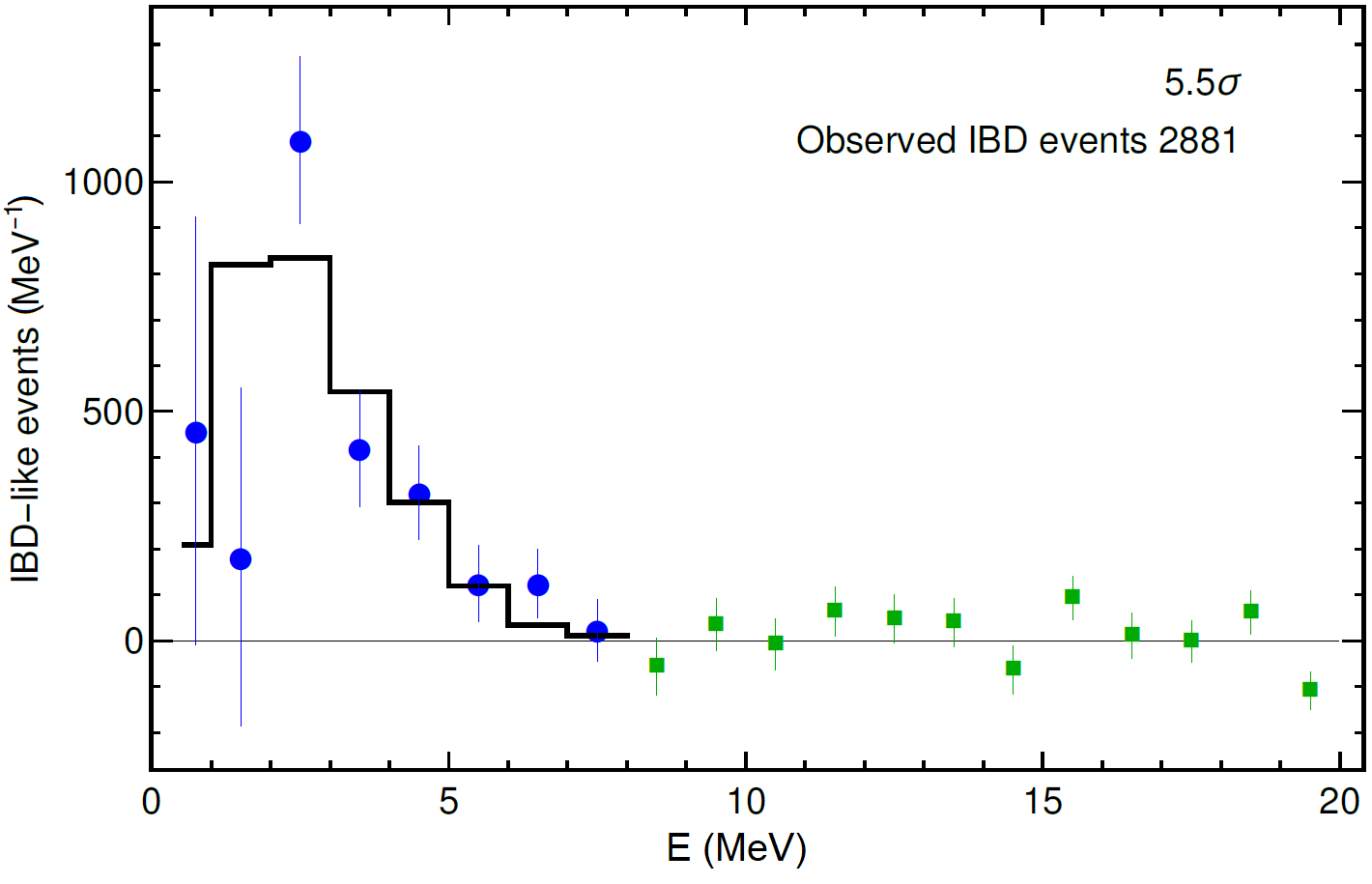}
\caption{\label{IBD} Shown is the difference between
  reactor-on correlated events and the reactor-off correlated
  events. The data points in green were used to determine the
  reactor-off normalization in this subtraction. The blue data points 
  are in the IBD-signal region and the histogram is the best-fit 
  Monte Carlo IBD spectrum.}
\end{figure*}

In a spectrum-based analysis we can exploit the fact that no IBD
events are expected above 8\,MeV\@ and thus we can use this part of 
the spectrum as a side band to normalize the reactor-on/off periods 
relative to each other.  We calculated a scale factor by taking the 
ratio of reactor-on and reactor-off correlated events in the 8 to 
20~MeV region.  Our computed scale factor is $1.666\pm0.013$, which 
turns out to be very close to the factor we get from a dead-reckoning 
of the relative reactor-on/reactor-off live-time, $1.673\pm0.005$.  
That we get this good agreement in spite of the $\sim$4\% RMS on the 
atmospheric pressure correction, is due to the fact that the difference 
between the average pressure corrections in the reactor-on and 
reactor-off periods is, by chance, quite low (less than 0.6\%).  This 
scale factor is applied to the correlated event numbers in all energy 
bins of the reactor-off spectrum, and then we perform the reactor-off 
subtraction.  The resulting spectrum is shown in Fig.~\ref{IBD}.  The 
error bars are obtained by propagating the error on the correlated 
event rate from the $\Delta$-t fit in each bin from both the 
reactor-on and reactor-off periods. The bin-to-bin correlated
error from the scale factor is not shown in the plot, but it is
included in the computation of the signal significance.

We perform a one-parameter fit of the observed reactor-on/off 
difference to the predicted signal spectrum.  In this fit we fully
account for the statistical uncertainty of the normalization between
the reactor-on and reactor-off data sets. The result of the fit is 
the best-fit value ($\hat a$) of the amplitude, $a$.  This is compared 
to the null hypothesis, where $a=0$, and the signal significance is
\begin{equation*}
    \sqrt{\chi^2(\hat a)-\chi^2(a\!=\!0)}.
\end{equation*}
Our best fit value corresponds to $2880\pm528$ IBD events,
for a ratio of observed to expected events of $82\% \pm 15\%$.  Given
that the distance cut has a simulated efficiency of $67\%$, and the
topological cuts are expected to be very efficient for true IBD
events, this is within expectations.  Overall, this constitutes a
5.5\,$\sigma$ detection of reactor neutrinos, in a detector with no
overburden. Our signal has the expected temporal, spatial and
energy signature expected for true IBD events.

Not surprisingly, we find the efficiency of the spectral analysis to
be more than twice that of the most sensitive rate analysis.  In the 
rate analysis, the low-energy bins, where the signal-to-noise is poor, 
can still contribute to the signal significance because their large 
uncertainties are contained bin-by-bin, such that they do not dilute 
the significance of the higher-energy bins where the signal-to-noise is 
much better.  Furthermore, the spectral analysis does not rely on sizable
corrections from the DAQ efficiency and atmospheric pressure, which makes
it inherently more robust.

\section{Outlook}

An 80\,kg prototype as presented in this paper is sufficient to
demonstrate reactor-on/off detection of a multi-gigawatt reactor over
a period of a few weeks. However, in a safeguards context there are
numerous ways in which this information can be obtained much more
easily without the recourse to neutrinos.

The unique capability offered by neutrino reactor monitoring is an {\it
in-situ}, quasi-real-time determination of the core inventory of
plutonium isotopes.  All use case scenarios of reactor neutrino monitoring
that go beyond a mere reactor-on/off detection, require a
high-statistics measurement of the neutrino spectrum.  Plutonium
production reactors typically have a thermal power in the
20-200\,MW$_\mathrm{th}$ range, thus requiring a fairly sizable active
detector mass. For instance, the case studies presented in
Refs.~\cite{Christensen:2013eza,Christensen:2014pva} are based on a
notional 5-ton, 100\% efficient detector. In order to stay within the
weight limits of a typical shipping container a detector module should
not exceed 20 metric tons, translating to a required overall neutrino
detection efficiency of about 25\%. Furthermore, while liquid scintillator
may not be a technical impossibility for safeguards applications, it
would require significant engineering controls to be practical, making
this technology easier to reject for a host country. In summary, the
results presented here,  establish a highly efficient liquid-free, 
unshielded detector with full spectral measurement capabilities.  In other 
words, a real step in the direction of a practical safeguards detector.

The MiniCHANDLER project was undertaken with the singular goal of
demonstrating the detection of reactor neutrinos and their energy
spectrum with this novel technology. Bench tests with our
MicroCHANDLER prototype have shown that the combination of new PMTs
(Hamamatsu R6321-100) and light guides improves the energy resolution
by a factor of two over the Amperex XP2202 PMTs alone, as implemented 
in this version of MiniCHANDLER.  Critically, the proposed new optics 
provides a clean resolution of the 511\,keV gamma's Compton edge, 
which will allow us to implement topological selections with greatly 
improved fast-neutron rejection efficiency.

Other future improvements include an upgrade of the electronics, based
on the SoLid detector readout~\cite{Arnold:2017lph}.  This will have
at least three known benefits: 1) increasing the dynamic range by a
factor of four, 2) fixing an undershoot/overshoot in the analog signal
affecting high primary-energy event pairs with $\Delta t<40\,\mu$s,
and 3) eliminating electronics cross talk.  Additionally, we will
double the $^6$Li concentration by putting a neutron sheet in the
middle of each cube layer.  Simulations show that this, so called
``half-cube'' modification should increase the $^6$Li capture
efficiency by 35\%, while decreasing the capture time by
48\%~\cite{haghighat}.  After returning from North Anna we tested this
configuration by modifying a single layer of the MiniCHANDLER
detector.  We found that it reduced the capture time and increased the
$^6$Li capture rate in agreement with simulation, while having no
measurable effect on the light collection.  Finally, simulations show
that adding just a meter of water equivalent shielding would reduce
the fast neutron background by an order of magnitude~\cite{haghighat}.
Future deployments of CHANDLER detectors will likely be accompanied by
a water tank, which can be filled on site, to provide an overburden of
up to one meter.

With the aforementioned improvements we expect to achieve a 
signal-to-noise ratio of better than one-to-one, which is essential for 
the safeguards goal of determining the plutonium  
content from distortions in the neutrino spectrum. \\

\acknowledgments This work was supported by the National Science
Foundation, under grant number PHY-1740247; the U.S. Department of
Energy Office of Science under award number DE-SC0018327; the U.S. Department of Energy
National Nuclear Security Administration Office of Defense Nuclear
Nonproliferation R\&D through the consortium for Monitoring,
Technology and Verification under Award number DE-NA0003920; Virginia
Tech's Institute for Critical Technology and Applied Science; Virginia
Tech's College of Science; the Office of the Vice President of
Research and Innovation at Virginia Tech; Virginia Tech's College of
Engineering; and Virginia Tech's Institute for Society, Culture and
Environment.  We are grateful for the cooperation and support of
Dominion Energy, and in particular the staff of the North Anna
Generating Station.

\bibliography{refs}

\end{document}